\documentclass[aps,prl,letter,twocolumn,dcolumn]{revtex4-1}
\usepackage{amsmath,amsthm}
\usepackage{amssymb,latexsym}
\usepackage{enumerate}
\usepackage[T1]{fontenc}
\usepackage{array}
\usepackage[all]{xypic}

\usepackage{tikz}
\usetikzlibrary{decorations.pathmorphing}
\usetikzlibrary{shapes,decorations.markings,decorations.pathreplacing,backgrounds,positioning,calc}

\def\CCC{{\mathbb{C}}}

\def\NN{{\mathbb{N}}}
\def\MM{{\mathbb{M}}}

\def\RR{{\mathbb{R}}}

\def\CB{{\cal B}}

\def\CH{{\cal H}}

\def\CL{{\cal L}}

\def\CR{{\cal R}}
\def\CS{{\cal S}}

\def\rem{\textnormal{rem}} 
 
\def\snull{\textnormal{ }}

\newcommand{\ra}{\rightarrow}

\def\ov{\overline}

\theoremstyle{definition}

\begin{document}
\preprint{APS/123-QED}
\title{Quantifier elimination theory and maps which preserve semipositivity}

\author{Grzegorz Pastuszak}
\email{past@mat.umk.pl, corresponding author}
\author{Adam Skowyrski}
\email{skowyr@mat.umk.pl}
\affiliation{Faculty of Mathematics and Computer Science, Nicolaus Copernicus University, Toru\'n, Poland}
\author{Andrzej Jamio{\l}kowski}
\email{jam@fizyka.umk.pl}
\affiliation{Faculty of Physics, Astronomy and Informatics, Nicolaus Copernicus University, Toru\'n, Poland}


\begin{abstract} We give an algorithm determining whether a hermiticity-preserving superoperator is positive. In our approach we apply techniques of quantifier elimination theory for real numbers. Furthermore, we argue that quantifier elimination theory should play more significant role in quantum information theory and other areas as well.  
\end{abstract}

\maketitle

The dynamics of a finite isolated quantum system is usually described by a one-parameter group of unitary transformations in a complex Hilbert space \cite{BZ}. However, in many physical problems it is necessary to consider a given quantum system as an open one which interacts with its surroundings. 

In modelling of open systems for which time behaviour can be represented by stochastic processes, one assumes that a system in question is described by certain mathematical model, for example by random variables in the classical case or by sets of non-commuting observables in the quantum case, acting on an abstract probability space. Correlations among subsystems of composite quantum systems, known as \textit{entanglements}, can be described in a rather ineffective manner. It appears that problems of the above type are naturally connected with some aspects of quantum dynamics and properties of dynamical maps. So-called \textit{entanglements witnesses} are related with the concept of \textit{positive maps}. 

In algebraic formulation of quantum mechanics, a fixed quantum mechanical system is represented by an algebra $A$ of operators acting on some Hilbert space $\CH$. In this approach, the observables (i.e. measured quantities) of the system are identified with hermitian (i.e. selfadjoint) elements in $A$ and physical states are given by the set $S(\CH)$ of density operators, that is, semipositive elements in $A$ with unital trace. Evolutions of the system are described by maps on the set $S(\CH)$. This means that we are interested in positive maps, that is, maps sending semipositive operators to semipositive operators. The general form of superoperators which preserve hermiticity of operators is well known \cite{HZ}. An important problem of finding among such superoperators those which preserve semipositivity is still an open one. In this paper we address the problem by applying techniques of quantifier elimination theory.

Quantifier elimination is a concept that appears in a field of mathematical logic called \textit{model theory} \cite{M,Rot}. It is especially important in the \textit{first-order logic} which is roughly the same as the \textit{predicate calculus}. Informally, quantifier elimination, if possible, allows to associate with a first-order formula $\varphi$ a quantifier-free formula $\varphi'$ in such a way that these two formulas are equivalent. Recall that a formula is \textit{quantifier-free} if and only if it does not contain quantifiers $\exists$ and $\forall$. Therefore quantifier-free formulas are \textit{computable conditions} which are straightforward to verify, unlike in the case of general formulas. In this sense, quantifier elimination can be viewed as a method for verifying validity of complicated formulas.  

As some straightforward examples of quantifier elimination, we can consider the well-known conditions for the existence of real roots of real quadratic, cubic and quartic polynomials, as well as concrete formulas for their complex roots. More generally, consider a field $K$ and a first-order formula $\varphi$ which states that two non-zero univariate polynomials $p,q\in K[x]$ have a common root in the algebraic closure of $K$. Then $\varphi$ is equivalent with the quantifier-free formula stating that the \textit{resultant} of $p$ and $q$ is non-zero \cite{Mi,Wa}. More advanced example is given by the \textit{quartic problem} which concerns finding conditions on real numbers $p,q,r$ so that $x^{4}+px^{2}+qx+r$ is a non-negative real number, for any $x\in\RR$. It is proved in \cite{AM} that this assertion holds if and only if $\delta\geq 0$ and $$p\geq 0\vee L<0\vee (L=0\wedge q=0)$$ where $L=8pr-9q^{2}-2p^{3}$ and $$\delta=256r^{3}-128p^{2}r^{2}+144pq^{2}r+16p^{4}r-27q^{4}-4p^{3}q^{2}.$$ We refer the reader to \cite{XY2} for similar considerations, see also \cite{Ar,Ch,Ya}.

The above examples are in fact instances of some of the most prominent results in model theory. The first one, known as the \textit{Tarski-Seidenberg theorem} \cite{Ta,vdD}, states that \textit{the theory of real closed fields admits quantifier elimination}. The second one, proved solely by A. Tarski \cite{Ta}, states that \textit{the theory of algebraically closed fields admits quantifier elimination}. The crucial consequence of these two theorems is that we are able to eliminate quantifiers in formulas (properly) composed from equalities and inequalities of real multivariate polynomials (in the case of Tarski-Seidenberg theorem), as well as from equalities of complex multivariate polynomials (in the case of Tarski's theorem). Importantly, in both cases this can be done in an \textit{effective way}, that is, we can compute a quantifier-free formula equivalent with the given one. This opens a possibility for applications of quantifier elimination theory in many areas of physics, applied mathematics and other fields which model their questions within mathematics. Indeed, assume that we are dealing with a \textit{scientific problem} which has the following general form: \begin{center}\textit{Determine whether some mathematical object $\omega$ possesses some property $\pi$.}\end{center} In many cases such assertions can be stated as first-order formulas over $\RR$ or $\CCC$. If this is indeed the case, we can compute a quantifier-free formula which is equivalent with the original statement. As argued above, this formula can be easily verified, so we get a complete solution to the problem we started with.

The above idea is realized in \cite{Pa} where we apply quantifier elimination theory for algebraically closed fields in order to find computable conditions for irreducibility of completely positive maps \cite{HZ,BZ}. Assume that $\Phi:\MM_{n}(\CCC)\ra\MM_{n}(\CCC)$ is completely positive, that is, there are matrices $K_{1},\dots,K_{s}\in\MM_{n}(\CCC)$ (called \textit{Kraus coefficients} of $\Phi$) such that $$\Phi(X)=\sum_{i=1}^{s}K_{i}XK_{i}^{*},$$ for any $X\in\MM_{n}(\CCC)$. The well-known result of D. Farenick \cite{Fa} states that $\Phi$ is irreducible if and only if its Kraus coefficients do not have a non-trivial \textit{common invariant subspace}. This means that if $V$ is a subspace of $\CCC^{n}$ such that $K_{i}V\subseteq V$, for any $i=1,\dots,s$, then $V=0$. We show in \cite{Pa} that this condition can be stated as some first-order sentence $\varphi$ over the field $\CCC$ of complex numbers. Moreover, we give a simple and straightforward proof of Tarski's theorem which is based on the effective version of Hilbert's Nullstellensatz \cite{J}. Then, applying this result, we compute the quantifier-free formula $\varphi'$ such that $\varphi$ is equivalent with $\varphi'$, and hence $\varphi'$ becomes the computable condition we are looking for. In this way we give a complete solution of the problem studied in \cite{JP1,JP2} and many other papers, see for example \cite{Sh,AI,AGI,GI,Ts,JKP1}. 

We stress that quantifier elimination for algebraically closed fields is generally much easier than the one for real closed fields, especially in terms of its proof and computational complexity \cite{H,Mi}. However, this Letter is devoted to show an application of quantifier elimination for real closed fields in quantum information theory. Indeed, we study the problem of determining whether a hermiticity-preserving superoperator is a positive map. Our strategy is the following. Assume that $\Phi:\MM_{n}(\CCC)\ra\MM_{n}(\CCC)$ is a superoperator which preserves hermiticity. We associate with $\Phi$ some real multivariate polynomial $p_{\Phi}$ in $4n$ variables such that $\Phi$ is positive if and only if $p_{\Phi}(a_{1},\dots,a_{4n})\geq 0$, for any $a_{1},\dots,a_{4n}\in\RR$ (this is denoted by $p_{\Phi}\geq 0$). In terms of first-order logic, the latter condition means that we consider the validity of the following first-order formula, say $\varphi$, over the field $\RR$ of real numbers: $\forall_{a_{1}}\forall_{a_{2}}\dots\forall_{a_{4n}\snull}p_{\Phi}(a_{1},\dots,a_{4n})\geq 0$. Note that the formula $\varphi$ has a special simple form - it is a negation of an existential formula. Therefore it is possible to apply the results of J. Renegar from \cite{Re1} which deals with the \textit{existential theory} of real numbers, see also \cite{Re2,Re3} for complete theory of quantifier elimination for the field $\RR$. In this way we obtain a procedure which determines whether the formula $\varphi$ holds or not. In particular, we are not interested in the explicit quantifier-free form $\varphi'$ of $\varphi$, but our way is completely sufficient for applications (and rather close to determining $\varphi'$).  

Although similar approaches to the problem we consider are known (see especially \cite{Ja1,Ja2}, \cite{SkZ} and \cite{Ch}), this Letter is the first paper presenting a concrete procedure which determines the validity of $\varphi$.

We stress that, to the best of our knowledge, the results of J. Renegar \cite{Re1,Re2,Re3} provide the most straightforward approach to the generally difficult quantifier elimination for the field of real numbers. Importantly, they are also quite effective from the point of view of computational complexity. However, it is the very nature of quantifier elimination theory for real closed fields that yields its algorithms are rather laborious. In some sense, this is a cost of the fact that these procedures can be applied to \textit{any given formula}. Thanks to this generality we are able to determine whether an \textit{arbitrary} hermiticity-preserving superoperator is positive or not. It is our opinion that such a goal cannot be achieved by any other methods. 

Details on computational complexity of quantifier elimination algorithms can be found in papers \cite{Re1,Re2,Re3} and monographs \cite{Mi,BPR}. We recommend the huge monograph \cite{BPR} for a comprehensive treatment of various aspects of quantifier elimination theory.    

Assume now that $\Phi\in\CL(\MM_{n}(\CCC))$ is a superoperator that preserves hermiticity. Our first goal is to introduce some real homogeneous polynomial $p_{\Phi}$ of degree $4$ in $4n$ variables such that $\Phi$ is positive if and only if $p_{\Phi}\geq 0$. Recall that there exists an isomorphism of Hilbert spaces $J:\CL(\MM_n(\CCC))\to\MM_n(\CCC)\otimes\MM_n(\CCC)$, known as the \textit{Choi-Jamio{\l}kowski isomorphism} \cite{Ja1,Ja2}, defined by the formula $$J(\Phi)=\sum_{i,j=1}^{n}E_{ij}\otimes\Phi(E_{ij}),$$ for any $\Phi\in\CL(\MM_n(\CCC))$. Here $E_{ij}$, for $i,j=1,\dots,n$, denotes the $n\times n$ complex matrix $[e_{kl}]$ such that $e_{kl}\in\{0,1\}$ and $e_{kl}=1$ if and only if $k=i$ and $l=j$. This isomorphism has the crucial property that a superoperator$\Phi$ is positive if and only if $J(\Phi)$ is block positive. Recall that an operator $$T\in\CL(\CCC^n\otimes\CCC^n)\cong\MM_n(\CCC)\otimes\MM_n(\CCC)$$ is \emph{block positive} if and only if $\langle x\otimes y\mid T(x\otimes y)\rangle$ is a non-negative real number, for any $x,y\in\CCC^{n}$. It is well-known that if $\Phi\in\CL(\MM_n(\CCC))$ preserves hermiticity, then $J(\Phi)$ is selfadjoint. Therefore we aim to solve a more general problem. Indeed, we shall express the block positivity of a selfadjoint operator $T$ as the condition $p_{T}\geq 0$ for some real multivariate polynomial $p_{T}$. 



Assume that $T\in\CL(\CCC^{n}\otimes\CCC^{n})$ is a selfadjoint operator. We denote by $T_{(ij)(kl)}$ the complex numbers such that $$T(\epsilon_{ij})=\sum_{k,l=1}^{n}T_{(ij)(kl)}\epsilon_{kl}$$ where $\epsilon_{ij}=e_i\otimes e_j$, for $i,j=1,\dots,n$, and $e_1,\dots,e_n$ are the elements of the standard $\CCC$-basis of $\CCC^{n}$. Since $T$ is selfadjoint, we get $\ov{T_{(ij)(kl)}}=T_{(kl)(ij)}$ and thus $T_{(ij)(ij)}\in\RR$, for any $i,j,k,l=1,\dots,n$. Assume that $x=(x_{1},\dots,x_{n}),y=(y_{1},\dots,y_{n})\in\CCC^{n}$. Then we have $x\otimes y=\sum_{i,j=1}^{n}x_{i}y_{j}\epsilon_{ij}$ which yields $$\langle x\otimes y \mid T(x\otimes y)\rangle=\sum_{i,j=1}^{n}\sum_{k,l=1}^{n}T_{(ij)(kl)}\ov{x_{k}y_{l}}x_{i}y_{j}.$$ It is clear that, for any $i,j=1,\dots,n$, the number $$\sigma_{(ij)}:=T_{(ij)(ij)}\ov{x_{i}y_{j}}x_{i}y_{j}$$ is a real number. Moreover, for any $i,j,k,l=1,\dots,n$, we have $$\ov{T_{(ij)(kl)}\ov{x_{k}y_{l}}x_{i}y_{j}}=T_{(kl)(ij)}\ov{x_{i}y_{j}}x_{k}y_{l},$$ so if $(ij)<(kl)$ in the lexicographical order, then the number $$\tau_{(ij)(kl)}:=T_{(ij)(kl)}\ov{x_{k}y_{l}}x_{i}y_{j}+T_{(kl)(ij)}\ov{x_{i}y_{j}}x_{k}y_{l}=$$$$=2Re(T_{(ij)(kl)}\ov{x_{k}y_{l}}x_{i}y_{j})$$ is also real. In fact, we may view both $\sigma_{(ij)}$ and $\tau_{(ij)(kl)}$ as real homogeneous polynomials of degree $4$ such that $$\sigma_{(ij)},\tau_{(ij)(kl)}\in\RR[x_{i}^{1},x_{i}^{2},x_{j}^{1},x_{j}^{2},y_{k}^{1},y_{k}^{2},y_{l}^{1},y_{l}^{2}]$$ where $x_{i}=x_{i}^{1}+x_{i}^{2}\iota$, $y_{i}=y_{i}^{1}+y_{i}^{2}\iota$, for any $i=1,\dots,n$, and $\iota$ is the imaginary unit. Therefore the polynomial $$p_{T}:=\sum_{i,j=1}^{n}\sigma_{(ij)}+\sum_{(ij)<(kl)}\tau_{(ij)(kl)}$$ is a homogeneous polynomial of degree 4 in $4n$ variables. We call $p_{T}$ the \textit{positivity polynomial} for $T$. The above arguments imply that a selfadjoint operator $T$ is block positive if and only if its positivity polynomial $p_{T}$ satisfies the condition $p_{T}\geq 0$. 

Assume that $\Phi(X)=\sum_{r=1}^{s}\alpha_{r}A_{r}XA_{r}^{*}$ and $A_{r}=[a_{ij}^{r}]$, for any $r=1,\dots,s$. Then $\Phi$ preserves hermiticity, so $J(\Phi)$ is selfadjoint and it follows that $\Phi$ is positive if and only if the positivity polynomial $p_{\Phi}:=p_{J(\Phi)}$ satisfies $p_{\Phi}\geq 0$. Observe that $$J(\Phi)_{(ij)(kl)}=e_{l}^{tr}\Phi(E_{ki})e_{j}=\sum_{r=1}^{s}\alpha_{r}a_{lk}^{r}\ov{a_{ji}^{r}}.$$ Then some straightforward calculations yield $$p_{\Phi}=p_{J(\Phi)}=\sum_{i,j=1}^{n}\sum_{k,l=1}^{n}J(\Phi)_{(ij)(kl)}\ov{x_{k}y_{l}}x_{i}y_{j}=$$$$=\sum_{i,j=1}^{n}\sum_{k,l=1}^{n}\sum_{r=1}^{s}\alpha_{r}a_{lk}^{r}\ov{a_{ji}^{r}}\ov{x_{k}y_{l}}x_{i}y_{j}=$$ $$=\sum_{r=1}^{s}\alpha_{r}\left\lVert[x_{1},\dots,x_{n}]\cdot A_{r}^{*}\cdot[y_{1},\dots,y_{n}]^{tr}\right\rVert^{2}.$$ This gives an interesting description of the positivity polynomial $p_{\Phi}$ as the sum of some non-positive or non-negative real multivariate polynomials, depending on signs of the numbers $\alpha_{1},\dots,\alpha_{s}$. As a consequence, we get an alternative proof of the fact that completely positive maps are positive. Indeed, if $\Phi$ is completely positive, then $\alpha_{1},\dots,\alpha_{s}>0$, so in this case it is obvious that $p_{\Phi}\geq 0$. This shows that the above description of $p_{\Phi}$ is useful. 

Now we show that the condition $p_{\Phi}\geq 0$ can be verified in an effective way. For this purpose, we apply some general techniques from the study of the existential theory of real numbers \cite{Re1}. These techniques get slightly simpler if we consider homogeneous polynomials of even degrees.

Assume that $g\in\RR[x_{1},\hdots,x_{n}]$ is a homogeneous polynomial of an even degree $d$. In Section 3 of \cite{Re1} the author constructs some finite set of real multivariate polynomials in variables $u_{1},\dots,u_{n},u_{n+1}$, depending on $g$. This set is crucial in our considerations, but we omit its construction, because it is too long and technical. We denote the set by $\CR_{g}$ in the Letter.

Assuming the set $\CR_{g}\subseteq\RR[u_{1},\dots,u_{n},u_{n+1}]$ is given, define $J(n,d):=\{0,\dots,nd^{2n}\}$ and $\CB(n,d)$ to be the set $$\{(i^{n-1},i^{n-2},\dots,i,1,0)\in\NN^{n+1}\mid i=0,\dots,nd^{2n}\}.$$ Let us fix $j\in J(n,d)$, $\beta\in\CB(n,d)$ and $r\in\CR_{g}$. We define some univariate polynomials $r_{1},\dots,r_{n+1}\in\RR[t]$ and $g_{j,\beta,r,+}^{un},g_{j,\beta,r,-}^{un}\in\RR[t]$ as follows: $$r_{i}(t)=(\frac{\partial r}{\partial u_{i}})(\beta+te_{n+1}),$$ for any $i=1,\dots,n+1$ and $$g_{j,\beta,r,+}^{un}(t)=g(r_{1}^{(j)}(t),r_{2}^{(j)}(t),\dots,r_{n}^{(j)}(t)),$$ $$g_{j,\beta,r,-}^{un}(t)=g(-r_{1}^{(j)}(t),-r_{2}^{(j)}(t),\dots,-r_{n}^{(j)}(t))$$ where $r_{i}^{(j)}$ denotes the $j$-th derivative of $r_{i}$. It follows from Section 4 of \cite{Re1} that $g\geq 0$ holds if and only if for any $j\in J(n,d)$, $\beta\in\CB(n,d)$ and $r\in\CR_{g}$ both conditions $$(\exists_{t}-g_{j,\beta,r,+}^{un}(t)>0\wedge r_{n+1}(t)>0)$$ and $$(\exists_{t}-g_{j,\beta,r,-}^{un}(t)>0\wedge -r_{n+1}(t)>0)$$ do not hold. We show that these conditions can be checked by applying the \textit{generalized Sturm's theorem}, also known as the \textit{Sturm-Tarski theorem}. In general, this theorem allows to calculate the number of distinct real roots of real univariate polynomials satisfying some additional conditions. 

In the Letter we apply the Sturm-Tarski theorem in determining the validity of the sentence $$\exists_{x\snull}(p(x)>0\wedge q(x)>0)$$ where $p,q\in\RR[x]$. In order to state the theorem, we recall the construction of the \textit{canonical Sturm sequence} \cite{KB,XY,BPR} (preceded by the construction of the \textit{generalized Sturm sequence}) associated with two non-zero polynomials $p,q\in\RR[x]$. Up to the sign, its elements are polynomials that occur as remainders in the Euclid's algorithm for determining the greatest common divisor of $p$ and $q$.

Assume that $p,q\in\RR[x]$ are non-zero polynomials. The generalized Sturm sequence is defined recursively in the following way. First we set $h_{0}=p$ and $h_{1}=q$. Assume that $n\geq 1$ and $h_{0},h_{1},\hdots,h_{n}$ are defined. If $h_{n}\mid h_{n-1}$, then the generalized Sturm sequence is the sequence $(h_{0},h_{1},\hdots,h_{n})$. Otherwise, we set $h_{n+1}=-\rem_{h_{n}}(h_{n-1})$ where $\rem_{h_{n}}(h_{n-1})$ is the remainder of division of polynomial $h_{n-1}$ by $h_{n}$. If $(h_{0},h_{1},\hdots,h_{n})$ is the generalized Sturm sequence for $p$ and $q$, then $h_{n}\mid h_{i}$, for any $i=0,\hdots,n$ and the sequence $$\left(\frac{h_{0}}{h_{n}},\frac{h_{1}}{h_{n}},\hdots,\frac{h_{n}}{h_{n}}=1\right)$$ is the canonical Sturm sequence.
 
Assume that $(h_{0},h_{1},\hdots,h_{n})$ is the canonical Sturm sequence for polynomials $p$ and $q$. We define $$\sigma_{a}(h_{1},\hdots,h_{s}):=(\sigma_{a}(h_{1}),\hdots,\sigma_{a}(h_{s}))\in\{+,-\}^{s}$$ where $a\in\{-\infty,\infty\}$ and $\sigma_{a}(h)=+$, if $\lim_{x\ra a}h(x)=\infty$ and $\sigma_{a}(h)=-$ otherwise, for any polynomial $h\in\RR[x]$. Define $\nu(p,q)$ as the number $$\lambda(\sigma_{-\infty}(h_{0},\hdots,h_{n}))-\lambda(\sigma_{\infty}(h_{0},\hdots,h_{n}))$$ where $\lambda(\alpha_{1},\hdots,\alpha_{s})$ denotes the number of all sign changes in $(\alpha_{1},\hdots,\alpha_{s})\in\{+,-\}^{s}$, that is, sequences of the form $(-,+)$ or $(+,-)$. 

For two non-zero $f,g\in\RR[x]$, denote by $N(f,g)$ the value of the number $$|\{x\in\RR\mid f(x)=0\wedge g(x)>0\}|$$ diminished by the number $$|\{x\in\RR\mid f(x)=0\wedge g(x)<0\}|$$ where $|X|$ is the number of elements of a finite set $X$. Observe that if $g$ is a polynomial such that $g>0$ (e.g. any positive constant polynomial), then $N(f,g)$ is the number of all distinct real roots of the polynomial $f$. 

The Sturm-Tarski theorem states that we have the equality $\nu(f,f'g)=N(f,g)$, and hence the number $N(f,g)$ can be computed. Observe that if $g=1$ is a constant polynomial, then the above theorem implies that the number $N(f,1)$ of all distinct real roots of $f$ can be computed as $\nu(f,f')$. This is the assertion of the original Sturm's theorem.

Now, assume that $f,p,q\in\RR[x]$ are non-zero polynomials and define $$\CS(f,p,q):=\{x\in\RR\mid f(x)=0\wedge p(x)>0\wedge q(x)>0\}.$$ Then \cite{KB} yields $|\CS(f,p,q)|$ is equal to the number $$\frac{1}{4}(N(f,p^{2}q^{2})+N(f,p^{2}q)+N(f,pq^{2})+N(f,pq)),$$ so in particular $|\CS(f,p,q)|$ can be computed. This gives us a way to determine whether the condition $$\exists_{x\snull}(p(x)>0\wedge q(x)>0),$$ say $\varphi$, holds. Indeed, first verify whether $$\lim_{x\ra a}p(x)=\lim_{x\ra a}q(x)=\infty$$ where $a=-\infty$ or $a=\infty$. If this is the case, then $\varphi$ holds. Otherwise, \cite{KB} or \cite{Re3} yields $\varphi$ is equivalent with the condition $$\exists_{x\snull}(p(x)>0\wedge q(x)>0\wedge (pq)'(x)=0).$$ The latter holds if and only if $|\CS((pq)',p,q)|\neq 0$.

Therefore our procedure for determining the positivity of a hermiticity-preserving superoperator $\Phi$ is complete. Indeed, for that purpose it suffices to calculate the positivity polynomial $p_{\Phi}$ and check whether $p_{\Phi}\geq 0$ by applying the methods described above. Although the procedure is rather time-consuming from the point of view of computational complexity, it seems to be the best possible, if the goal is to obtain a general solution to the problem we consider.

The authors are indebted to J. Renegar for his assistance in understanding the contents of \cite{Re1}.

\end{document}